\documentclass[journal]{IEEEtran}
\usepackage{epsfig}
\usepackage{hyperref}
\usepackage{amsmath}
\usepackage{dcolumn}
\usepackage{graphics}
\usepackage{graphicx}
\usepackage{float}
\usepackage{booktabs}

\usepackage{graphicx}
\usepackage{cite}
\usepackage{picinpar}
\usepackage{amsmath}
\usepackage{url}
\usepackage{lipsum}
\usepackage{flushend}
\usepackage[latin1]{inputenc}
\usepackage{colortbl}
\usepackage{soul}
\usepackage{multirow}
\usepackage{pifont}
\usepackage{color}
\usepackage{float}
\usepackage{rotating}
\usepackage{adjustbox}
\usepackage{lscape}
\usepackage{cite}
\usepackage{alltt}
\usepackage{enumerate}
\usepackage{siunitx}
\usepackage{epstopdf}
\usepackage{pbox}

\ifCLASSINFOpdf
\else
\fi
\begin{document}
%
\title{Machine Intelligent Techniques for Ramp Event Prediction in Offshore and Onshore Wind Farms}
%
%
%

\author{Harsh S. Dhiman,~\IEEEmembership{Student Member}
        Dipankar Deb,~\IEEEmembership{Senior Member}
\thanks{H. Dhiman is with the Department
of Electrical Engineering, Adani Institute of Infrastructure Engineering, Ahmedabad, 382421 India e-mail: harsh.dhiman@aii.ac.in}
\thanks{D. Deb is with Department
of Electrical Engineering, Institute of Infrastructure Technology Research and Management, Ahmedabad}}
\maketitle

\begin{abstract}
Globally, wind energy has lessened the burden on conventional fossil fuel based power generation. Wind resource assessment for onshore and offshore wind farms aids in accurate forecasting and analyzing nature of ramp events. From an industrial point of view, a large ramp event in a short time duration is likely to cause damage to the wind farm connected to the utility grid. In this manuscript, ramp events are predicted using hybrid machine intelligent techniques such as Support vector regression (SVR) and its variants, random forest regression and gradient boosted machines for onshore and offshore wind farm sites. Wavelet transform based signal processing technique is used to extract features from wind speed. Results reveal that SVR based prediction models gives the best forecasting performance out of all models. In addition, gradient boosted machines (GBM) predicts ramp events closer to Twin support vector regression (TSVR) model. Furthermore, the randomness in ramp power is evaluated for onshore and offshore wind farms by calculating log energy entropy of features obtained from wavelet decomposition and empirical model decomposition.
\end{abstract}


%
\IEEEpeerreviewmaketitle

\section{Introduction}
\IEEEPARstart{W}{ith} wind energy being a driving force, both, onshore and offshore technologies have attracted big investments globally \cite{2016}. Onshore wind farms have an advantage of proximity to the utility grid whereas offshore wind farms need long transmission cables to transmit power from the sea to grid. However, with a strong wind field in the offshore areas compared to onshore, the offshore wind farm installations have risen drastically. Further, with lesser turbulence and more uniform wind speed, the lifetime of wind turbines in offshore scenario increases \cite{Esteban2011}. Offshore wind farms have a disadvantage of higher installation and engineering costs accompanied by the wear and tear of generator and mechanical equipment. Onshore wind farms, on the other hand face the challenge in their acceptance from limited land availability, visual intrusion and damage to wildlife. The defining factor of operation between onshore and offshore wind farms is the surface roughness length. It is observed that, larger the surface length higher the reduction in wind velocity and theoretically the wind speed at ground is zero \cite{Aerodyna75:online}.  

Wind energy related installation activity, particularly in the offshore areas, is challenged by the transportation costs owing to large blades and support structures to be placed in deep sea water. Another issue is the uncontrolled vibrations arising from the combination of wind and wave force \cite{Colwell2009}. With a higher wind speed in the sea, the probability of ramp instances increases. Wind power growth has led to detailed study of ramp events, particularly categorized as ramp-up and ramp-down events with each characterized by sudden wind speed change in a short period. Mathematically, a wind power ramp event \cite{Kamath2010} is described as the arithmetic difference between wind power at consecutive time instants:
\begin{eqnarray}
\label{e1}
\Delta P_w^{Ramp}= P_w(t+\Delta t)-P_w(t).
\end{eqnarray} 

The occurrence of a wind power ramp event is ascertained by a threshold value of the total wind farm power. Wind power forecasting is an essential market procedure to regulate the electricity markets where the farm operators hold an advantage to derive profit from their generation given an accurate wind forecasting strategy. Predicting these ramp events can suitably prevent intermittent grid failures and for that purpose, proper characterization of ramp events is imperative in avoiding potential risks in the management of a power system. On the contrary, a poor characterisation of the ramp performance inevitably hinders the identification of the processes that cause ramp events.

In Horns Rev wind farm, large variation in wind power is seen in a short duration \cite{Vincent2010}. In a study carried out by Gjerstad et al., the variation in wind power is triggered by atmospheric properties of boundary layer \cite{Gjerstad1995}. The atmospheric factor that influences the wind speed variations is the mean lapse rate ($\partial T/\partial z$), that is, rate of change of air temperature ($T$) with height $z$. In case of offshore wind farms connected to German power grid, within 8 hrs, the wind power transfer between converters increased from 4 GW to 19 GW as a result of a sudden change in wind speed \cite{Kramer2014}.

Further, Nissen studies seasonal variations in wind speed over coastal areas of H{\o}vs{\o}re which shows that the wind speed variation are primarily dominant in the spring and winter season \cite{8220775345684ab4a6201dff3283f31f}. 
It is found that the wind power ramp events are characterized based on magnitude error, phase error and location error as pointed out by Potter et al.\cite{Potter2009}. In \cite{Cutler2007}, authors discuss the MesoLAPS and Wind power prediction tool (WPPT) for analyzing and categorizing large wind power ramp events where the problem is modeled as a classification task. Root mean squared error (RMSE) is evaluated over 1 year for two wind speed time series with 5-min and 10-min sampling intervals. Results reveal that the RMSE values are found to be more optimistic than conventional persistence and climatology methods. 

Gallego-Castillo et al. reviewed various wind ramp events and have described preliminary definition of a ramp event and the threshold values that various literature have used to formulate a ramp event \cite{GallegoCastillo2015}. Bossavy et al. proposed a reliable forecast methodology to characterize ramps with a derivative filtering edge detection approach and numerical weather prediction ensembles to make probabilistic forecasts of ramp occurrence \cite{Bossavy2012}. Understanding the underlying role of the synoptic weather regimes in triggering the wind power ramp events can help improve and complement the current forecast techniques. Cuoto et al. have identified and classified the weather regimes over mainland Portugal associated with the occurrence of severe wind power ramps \cite{Couto2015}.

Cornejo-Bueno et al. have described the machine learning techniques like, support vector regression (SVR), gaussian process regression (GPR), multi-layer perceptrons (MLP) and extreme learning machines (ELM) to forecast wind power ramp events \cite{CornejoBueno2017}. Cui et al. detect ramp event probability characteristics from scenarios captured bands to evaluate the ramp event forecasting method using a modified genetic algorithm with multi-objective fitness functions \cite{Cui2015}, \cite{Cui2015b}. Gallego et al, formulated a ramp function to describe a ramp event characterised by high power output gradients evaluated under different time scales, and is based on discrete wavelet transform to provide a continuous index related to the ramp intensity \cite{Gallego2014}. In another work, Cornejo-Bueno et al. have considered a neural network based model in tandem with evolutionary algorithm to optimize the parameters of several classifiers such as support vector machines (SVM) and extreme learning machine (ELM) \cite{CornejoBueno2018, Du2019, Zhang2015c}. Three wind speed datasets from Spain are considered to evaluate the accuracy of hybrid method for classifying wind power ramp events. In \cite{Li2017}, the authors have presented a feature extraction method based on Gabor filtering where the atmospheric pressure fields are taken into consideration. Results are compared with   state-of-the-art neural network method and it is found that Gabor method with its change in power as its output gives a better prediction performance.

In a recent study, Ouyang et al. have presented an improved ramp prediction methodology using residual correction through a model that combines the advantages of the auto-regression and Markov chain models \cite{Ouyang2019}. Classification based forecast is inaccurate and causes class imbalance issue in the machine learning problem since the occurrence of a ramp event is typically rare, and so Takahashi et al. have used and experimentally evaluated this sampling approach on a real-world wind power generation dataset \cite{Takahashi2017}. In another work, Fujimoto et al. have presented an altering scheme of an impending ramp event for efficient operational decisions using supervised learning algorithms for better health management and reduced downtime \cite{Fujimoto2019}. Zhang et al. decompose wind speed data and analyze each part using different models to make predictions based on wind speed ramp and residual distribution \cite{Zhang19}. In other ramp event prediction related works, Dhiman et al. have discussed SVR and its variants \cite{Dhiman2018} for predicting wind power ramp events where five datasets are tested for ramp-up and ramp-down events \cite{Dhiman2019}, \cite{harshels}.

The current work addresses following objectives:
\begin{enumerate}
	\item {Ramp prediction models for onshore and offshore wind farms are discussed. The wind powers corresponding to the wind speed time series obtained are calculated for identifying the ramp events. The wind power ramp events are predicted using a hybrid method involving wavelet transform decomposition and machine learning methods. Thus, the underlying problem here is modeled as a regression task with $\Delta t=$ 10 min.}
	\item {The potential capability of random forest regression and gradient boosted machines are checked. Several error metrics for the entire wind speed time series and absolute error for ramp events (ramp-up and ramp-down) are evaluated and further compared with benchmark persistence and SVR models.}
	\item {Log energy entropy based randomness is discussed for onshore and offshore wind farm sites. The randomness in a ramp signal is an important feature to be dealt with. Higher order randomness calls for accurate wind resource assessment and micro-siting.}
\end{enumerate} 

This manuscript is organized as follows. Section 2 describes machine learning methods, that is, support vector regression and its variants, random forest regression and gradient boosted machines followed by framework of ramp prediction models in Section 3. In Section 4 outcomes of the proposed models are presented and Discussions are highlighted in Section 5. Section 6 discusses conclusions.

\section{Methods for Wind Power Ramp Prediction}
Next, we discuss various methods employed to predict ramp events in onshore and offshore wind farms. Wavelet decomposition based machine learning methods are used. A hybrid model based on  Wavelet decomposition serves the purpose of eliminating stochastic trends in wind speed time series. Support vector regression and its variants like $\varepsilon$-SVR, Least square support vector regression, Twin support vector regression and $\varepsilon$-Twin support vector regression are discussed. Further, prediction models based on regression trees like Random forest and Gradient boosted machines are also discussed for predicting ramp events. 

\subsection{Support Vector Regression \& its variants}
Support vector regression (SVR) which is coined from support vector machines (SVMs) \cite{Vapnik2000} is a popular machine learning method used in several branches of forecasting like solar radiation forecasting \cite{Gala2013}, wind forecasting \cite{LIU2018} and hydrological time series \cite{Sahoo2018}. The basic idea of SVR is to map a set of non-linear features to a higher dimension where the data is linearly separable,  mathematically expressed as 
\begin{equation}
g(x)=w^Tx+b,
\end{equation}
where $w$ is a weight vector, $x$ represents set of input features and $b$ is the bias term. SVR operates on historical data which plays an important role in predicting the future values of the response variable. Consider a set $X\in R^n$, $X=[x_1, x_2,\dots,x_n]$ as the set of input features. The weights for the input features are obtained by solving an optimization problem described in Table \ref{T}. Further, Suykens penned Least square support vector regression (LSSVR), where the objective function for minimization incorporates the squared error term \cite{Suykens1999}. The computation time for LSSVR is fast compared to classical SVR owing to its smaller sized lagrangian multiplier matrix. In 2010, Peng derived Twin support vector regression (TSVR) where two smaller sized optimization problems are solved to arrive at resultant regressor \cite{Peng2010}. The optimization problem along with resultant regressor $g(x)$ is depicted in Table \ref{T}. Further, in 2012, Shao et al. presented a $\varepsilon$-Twin support vector regression that incorporates an extra regularization factor and arrives at the final regressor using successive over relaxation technique which is a fast convergence algorithm \cite{Shao2012}.  
However, with larger datasets, the regression analysis can be computationally expensive and can lead lead to large errors in predicting ramp events. Thus, the ramp event prediction is extended for two more methods, that is, random forest regression (RFR) and gradient boosted machines (GBM) categorically for onshore and offshore wind farm sites, and specifically for offshore wind farms, the dominating strong wind field poses a forecasting challenge for operators. 

	\begin{table}[h]
		\begin{center}
			\caption{$\varepsilon$-SVR, LS-SVR, TSVR and $\varepsilon$-TSVR regression models} \vspace{0.2cm}\label{T}
			\begin{tabular}{|c|}\hline
				Optimization Problem  \\ \hline
				Model: $\varepsilon$-SVR,~~ Regressor: $g(x)=w^{T}x+b$ \\
				$\text{min} \hspace{0.1cm} \frac{1}{2} \parallel w\parallel^2 + C(e^T\xi+e^T\xi^*)$ \\ 
				$ \textbf{s.t.}~~ y-w^{T}x-eb \le e\varepsilon +\xi, \xi \ge 0,$ \\
				$w^{T}x+eb-y \le e\varepsilon +\xi^{*}, \xi^* \ge 0$, \\ 
				\hline
				Model: LS-SVR\\
				Regressor: $g(x)=\sum_{i=1}^{n}\alpha_i k(x,x_i)+b$\\
				$\text{min} \hspace{0.2cm} \frac{1}{2} \parallel w \parallel^2 + \frac{1}{2}\gamma\sum_{i=1}^{n}\varepsilon_i^2$\\
				$\textbf{s.t.} \hspace{0.2cm} y_i= w^T\phi(x_i)+b+\varepsilon_i$ \\ \hline
				Model: TSVR\\
				Regressor: $g(x)=\frac{1}{2}((w_1+w_2)^Tx+(b_1+b_2))$\\
				$\text{min} \hspace{0.2cm} \frac{1}{2}\parallel(y-e\varepsilon_1-(xw_1+eb_1))\parallel^2+ C_1e^T\sum_{i=1}^{n}\xi_i$ \\
				$\text{min} \hspace{0.2cm} \frac{1}{2}\parallel(y-e\varepsilon_2-(xw_1+eb_2))\parallel^2+ C_2e^T\sum_{i=1}^{n}\eta_i$ \\
				$\textbf{s.t.} \hspace{0.2cm} y_i-(x_iw_1+eb_1)\ge e\varepsilon_1-\xi_i$ \\
				$\textbf{s.t.} \hspace{0.2cm} (x_iw_2+eb_2)-y_i\ge e\varepsilon_2-\eta_i$ \\ \hline
				Model: $\varepsilon$-TSVR\\
				Regressor: $g(x)=\frac{1}{2}((w_1+w_2)^Tx+(b_1+b_2))$ \\
				$\text{min} \hspace{0.2cm} \frac{1}{2}C_3(w_1^Tw_1+b_1^2)+\frac{1}{2}\xi^{*T}\xi+C_1e^T\xi$ \\
				$\text{min} \hspace{0.2cm} \frac{1}{2}C_4(w_2^Tw_2+b_2^2)+\frac{1}{2}\xi^{*^T}\xi+C_2e^T\eta$ \\
				$\textbf{s.t.} \hspace{0.2cm} Y-(Xw_1+eb_1)=\xi^*$
				,\\ $Y-(Xw_1+eb_1)\ge -e\varepsilon_1-\xi, \xi \ge 0$ \\
				$\textbf{s.t.} \hspace{0.2cm} Y-(Xw_2+eb_2)=\eta^*$
				 , \\ $Y-(Xw_2+eb_2)\ge -e\varepsilon_2-\eta, \eta \ge 0$ \\ 
				\hline
				Model: Kernel, $\varepsilon$-TSVR\\
				Regressor: $g(x)=\frac{1}{2}((w_1+w_2)^TK(X,X^T)+(b_1+b_2))$  \\
				$\text{min} \hspace{0.2cm} \frac{1}{2}C_3(w_1^Tw_1+b_1^2)+\frac{1}{2}\xi^{*T}\xi+C_1e^T\xi$ \\
				$\text{min} \hspace{0.2cm} \frac{1}{2}C_4(w_2^Tw_2+b_2^2)+\frac{1}{2}\xi^{*^T}\xi+C_2e^T\eta$ \\
				$\textbf{s.t.} \hspace{0.2cm} Y-(K(X,X^T)w_1+eb_1)=\xi^*$, \\ 
				$Y-(K(X,X^T)w_1+eb_1)\ge -e\varepsilon_1-\xi, \xi \ge 0$\\
				$\textbf{s.t.} \hspace{0.2cm} Y-(K(X,X^T)w_2+eb_2)=\eta^*$,\\ 
				$Y-(K(X,X^T)w_2+eb_2)\ge -e\varepsilon_2-\eta, \eta \ge 0$ \\ 
				\hline
			\end{tabular}
			\vspace{-0.25cm}
		\end{center}
	\end{table}


\subsection{Random forest regression}
Proposed by Brieman, random forest is an ensemble method that generates something, akin to a forest of trees from a given training sample \cite{Breiman2017}. The ensemble based models are far more accurate than a single method owing to advantages like capturing linearity and non-linearity of time-series obtained from individual methods. Random forest begins with splitting the input features into a group of subsets that essentially form a tree. Similar to hyperparameters ($\sigma$ and $C$) in SVR, in random forest regression, the number of trees and number of random features in each tree decomposition are the parameters that decide the performance of regression. At each decision tree, a fitting function is created which acts on the random features selected. Finally, at the end of the training process, a random forest model is created. It is worthwhile to note that during training process, each tree is created from randomly selected input vectors and thus it is called `random' forest. The estimated output of a random forest regression is given as
\begin{equation}
\hat{A}=\frac{1}{k}\sum_{i=1}^{k}\hat{r}\Big(X,V_i\Big),
\end{equation}   
where $\hat{r}\Big(X,V\Big)$ is the representative tree at the end of training process, $X$ is the set of input feature vectors and $T$ is the collective set representing input-output pair $V_i=(x_1,y_1),(x_2,y_2),\dots(x_n,y_n)$. 

\begin{figure}[h]
	\centering
	\includegraphics[width=1\linewidth]{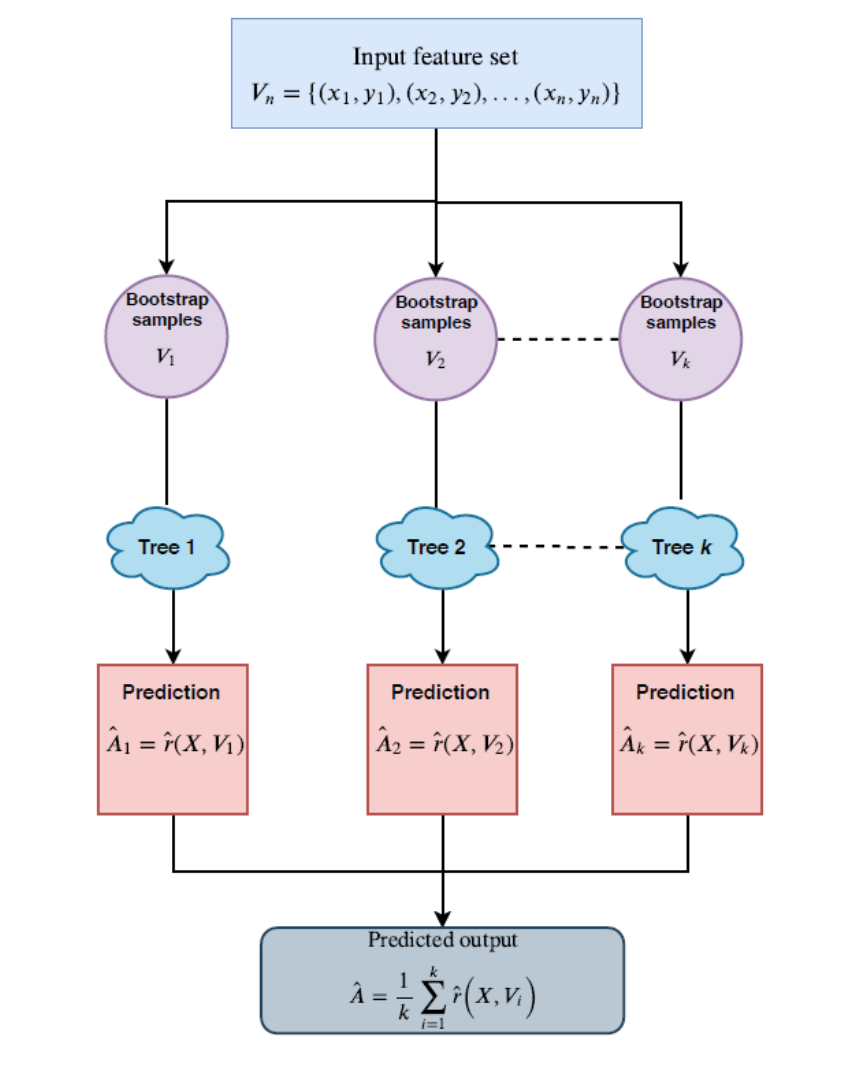}
	\vspace{-0.25cm}
	\caption{Block diagram for random forest regression model}\vspace{-0.15cm}
	\label{rfr_wavelet}
\end{figure}

A particular tree is characterized by a node that leads to number of branches, as depicted in Fig. \ref{rfr_wavelet}.

\begin{figure*}
	\centering
	\includegraphics[width=0.9\linewidth]{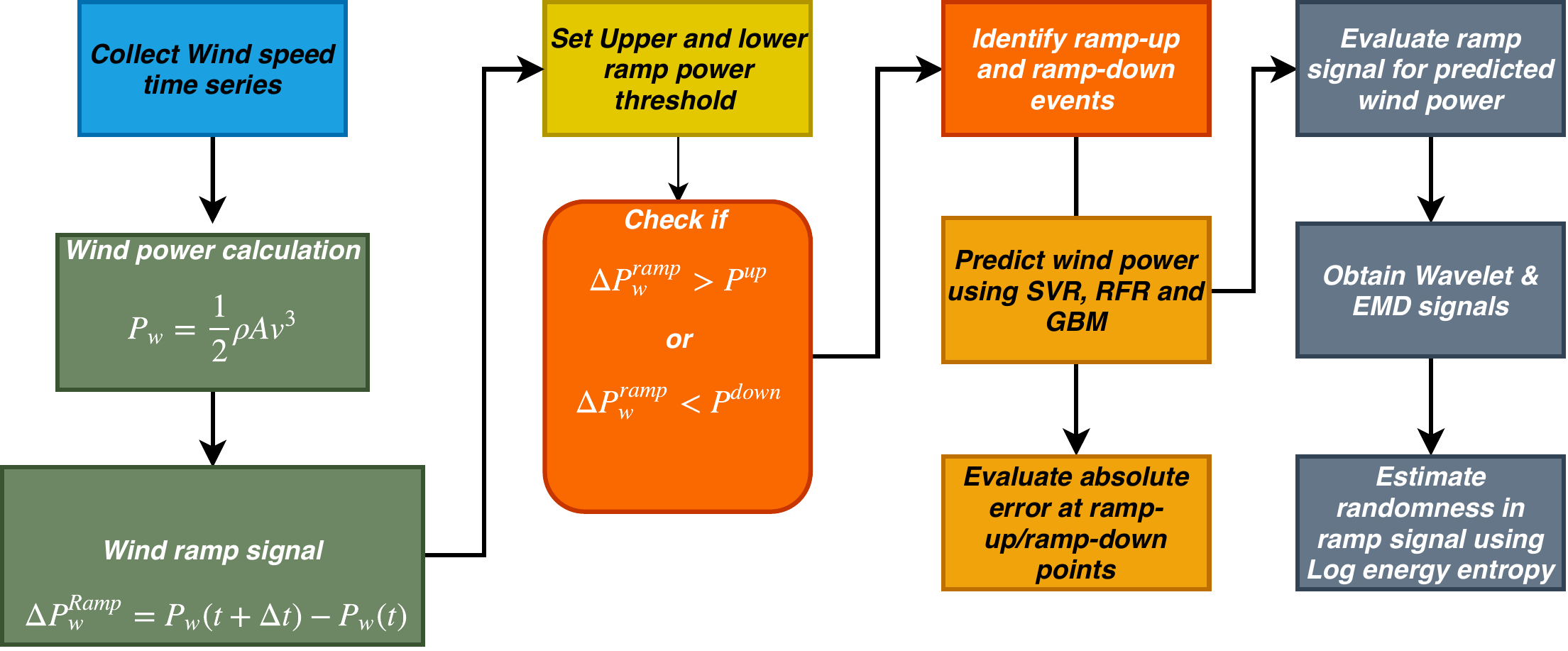}
	\caption{Block diagram for ramp event prediction and randomness in ramp signal}
	\label{fd}
\end{figure*}

The predicted output is averaged over $k$ decision trees. An added advantage of random forest regression is its insensitivity to noise due to uncorrelated trees via differential sampling of inputs. Figure \ref{fd} illustrates a generic block diagram of the wavelet based ramp prediction model for onshore and offshore wind farms. 

The input feature set or predictors are a set of approximation signal (A5) and detail signals (D1, D2, D3, D4 and D5) obtained from wavelet decomposition of wind speed. A common problem of over-fitting persists in machine learning regression models when a well trained model captures the noise component as well. To reduce the complexities posed by over-fitting, random forest makes a compromise between a flexible and an inflexible model. In the training phase, each regression tree draws a sample from the feature set and is drawn repeatedly. This ensures that even though the tree might possess a high variance but the overall variance of forest will be low. Random forest work on the principle of bagging that combines the predictions from different tree models to give an overall insight to the data under training. This also helps to reduce the potential over-fitting caused by supervised machine learning models. 

\subsection{Gradient boosted machines }
Gradient boosted machines (GBM) is an ensemble based regression method that takes into account the loss of the previously fit decision tree. In a GBM model, various weak learners combine together to arrive at an accurate model. Boosting based regression trees are effective in predicting accurate response values. A particular boosting algorithm strengthens the tree model by minimizing the inaccuracies of weak models in form of loss function \cite{Solomatine}. The main task is to reduce the error posed by weak learners in each iteration. The final predicted response is the linear combination of fit trees. The learning rate of each decision tree determines the predictive performance. A higher learning rate indicates less number of trees required for prediction and vice versa \cite{Elith2008}. Based on a cross-validation experiment the ideal number of trees required along with learning rate is obtained.

The aim for a individual model is to reduce the error generated by a loss function. Commonly used loss function is $\mathcal{L}_2$ function that minimizes the sum of squared errors between predicted and actual value. Consider $k$ such weak models that account for prediction of a variable $y$ given feature set $x=[x_1,\dots,x_n]$, mathematically it can be expressed as 
$\hat{y}=\sum_{i=1}^{k}f_i(x),$
where $f_i(x)$ represents each weak learner that is collectively trained to improve the prediction. The loss function is given as\vspace{-0.15cm}
\begin{equation}
    \mathcal{L}_2 =\frac{1}{N}L(y_i,\hat{y}_i),~~
    L(y_i,\hat{y}_i) =\sum_{i=1}^{N}(y_i-\hat{y}_i)^2,
\end{equation}
where $L(y_i,\hat{y}_i)$ is a loss function based on squared errors for $N$ observations. The aim of a GBM regression technique is to minimize the $\mathcal{L}_2$ loss function. However, the $\mathcal{L}_2$ loss function is more sensitive to outliers and can reduce the robustness of the model. For optimizing the hyper-parameters in GBM, a gradient descent algorithm is used which minimizes the cost function taking into account the negative of the gradient. Let us consider a $\mathcal{L}_2$ loss function, the gradient of this function with respect to predicted sample $\hat{y}_i$ is given as 
\begin{equation}
    \frac{\partial L(y,\hat{y})}{\partial \hat{y}_i}=\frac{\partial}{\partial \hat{y}_i}\sum_{i=1}^{N}(y_i-\hat{y}_i)^2 = -2(y_i-\hat{y}_i).
\end{equation}
The gradient reflects that while tracking the minimum point of the loss function, the GBM actually tracks the residual vector $(y-\hat{y})$. Similarly, for $\mathcal{L}_1$, the MAE can be tracked by finding the gradient of loss function. The essence of a GBM model lies in the boosting technique that combines all the weak models in a stage-wise manner than in a parallel manner as seen in random forest model. The GBM model in recursive form is given as $F_m(x)=F_{m-1}(x)+\eta\Delta_m(x)$,
where $\eta$ represents the learning rate and $\Delta_m$ refers to a regression model fitted to the residuals.

\begin{figure}[h]\vspace{-0.25cm}
\centering
   \includegraphics[width=0.99\linewidth]{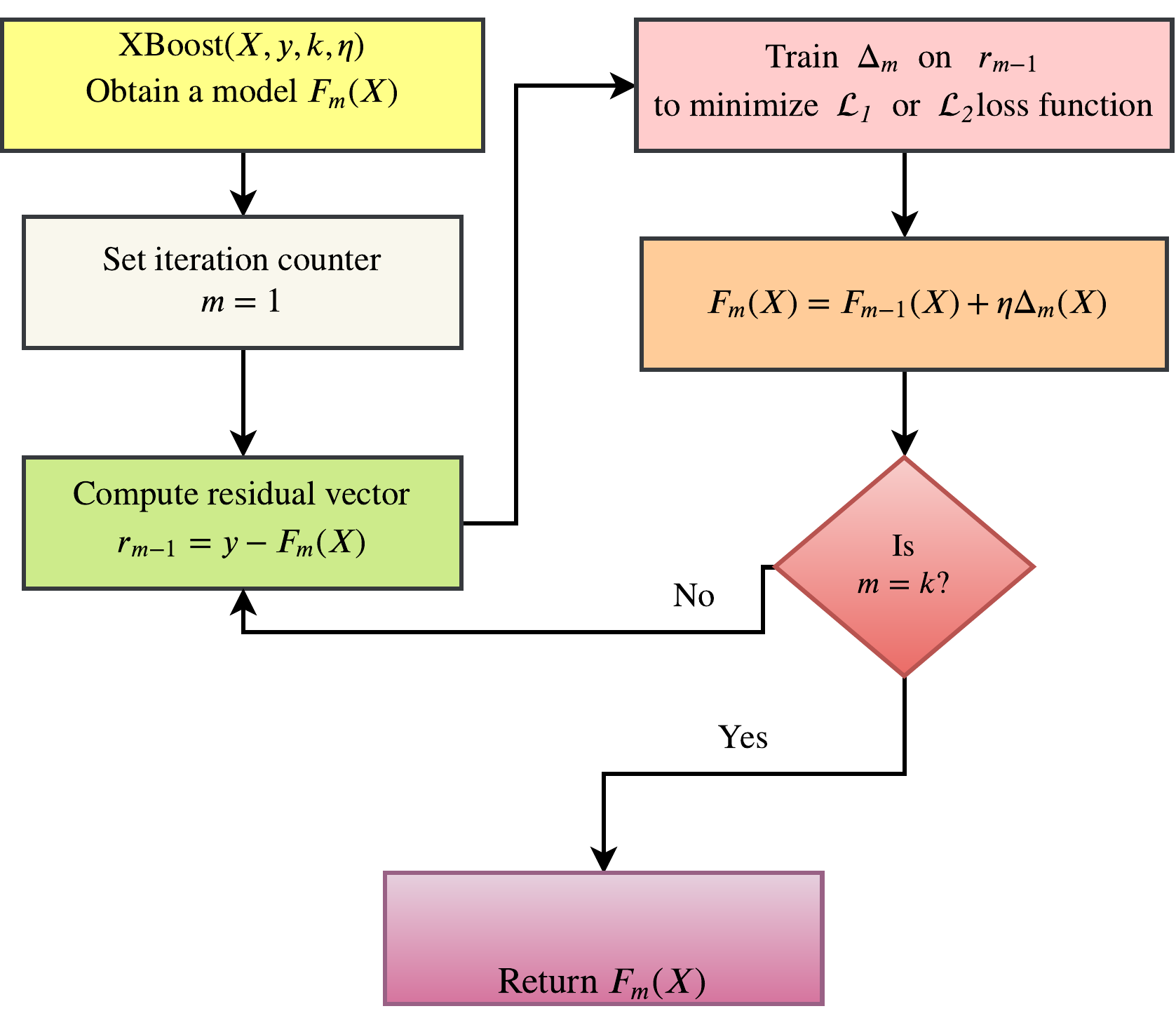}\vspace{-0.25cm}
   \caption{XBoost algorithm for Gradient boosted machines}\vspace{-0.25cm}
    \label{fig:XGBM}
\end{figure}

Figure \ref{fig:XGBM} illustrates the \textit{XBoost} algorithm used for obtaining the best regressor.  

\section{Datasets \& Framework of prediction models}
Wind speed variation depends on the terrain under study. For all the wind farms operating in neutral atmospheric boundary layer, the wind speed follows a logarithmic profile with height. For onshore wind farms, the surface roughness length ($z_0$), is around 0.005 m while for offshore wind farms it is 0.0002 m. The current hybrid prediction model is based on a combination of wavelet transform and a machine learning algorithm. Wavelet transform is widely used for defragmentation of a time-series signal into low and high frequency sub-signals called as approximate and detail signals respectively. A 5-level db4 wavelet transform is used to decompose the wind speed time series. The approximate signal (A5) along with detail signals (D1, D2, \dots, D5) form input feature set for all the prediction models. The datasets collected for the month of March 2019 are wind speed time-series at a height of 10 m above ground averaged over a time interval of 10 minutes. For an accurate wind power ramp event study, it is desirable to transform the available wind speed to a hub height of 90 m using logarithmic law \cite{Irwin1979}. In this work, we consider all the wind farms with identical wind turbines having rotor diameter of 120 m with a hub height of 90m. The datasets (onshore and offshore) with their site coordinates and statistical parameters are depicted in Table \ref{TCR7}. Mean represents the mean of the wind speed time-series and SD indicates the standard deviation. The variability in wind speed is high as the changes in wind speed on minute to minute scale are likely to cause large power changes.

\begin{table}[h]
	\caption{Description of wind farm datasets for the month March 2019}\label{TCR7}\vspace{0.2cm}
	\centering
	\begin{tabular}{|c|c|c|}\hline
		Onshore Wind farm  & Mean & SD  
		\\
		(Site coordinates) & &
		\\	
		\hline
		Amakhala Emoyeni, SA (a) & 6.264 & 3.198 \\
		($-32.17^\circ$ N, $25.95^\circ$ E) & &\\
		Clyde, Scotland (b) & 3.829 & 1.626\\
		($55.46^\circ$ N,  $03.65^\circ$ W) & &\\
		Gansu, China (c) & 4.000 &2.477\\
		($40.20^\circ$ N, $96.90^\circ$ E) & &\\
		McCain Foods, UK (d) & 6.491 & 3.519\\
		($52.56^\circ$ N, $0.172^\circ$ W) & & \\
		Shephards Flat, USA (e) & 6.074 & 2.618\\
		($45.70^\circ$ N, $120.06^\circ$ W) & &\\
		Akhfenir, Morocco (f) & 3.096 & 1.505\\
		($27.95^\circ$ N, $12.00^\circ$ W) & &\\
		\hline 
		Offshore Wind farm & Mean & SD  \\	
		(Site coordinates) & &\\
		\hline
		Amrumbank, Germany (A) & 11.176 & 4.962 \\
		($54.50^\circ$ N, $7.80^\circ$ E) & &\\
		Anholt, Denmark (B) & 8.999 & 3.537\\
		($56.60^\circ$ N, $11.21^\circ$ E) & &\\
		Gemini, Netherlands (C) & 7.577 &4.174\\
		($54.03^\circ$ N, $5.96^\circ$ E) & &\\
		HornsRev 2, Denmark (D) & 11.183 & 4.546\\ 
		($55.60^\circ$ N, $7.59^\circ$ E) & &\\
		Veja Mate, Germany (E) & 11.490 & 4.685\\
		($54.31^\circ$ N, $5.87^\circ$ E) & &\\
		Walney, UK (F) & 11.342 & 5.015\\
		($54.04^\circ$ N, $5.92^\circ$ W) & &\\
		\hline 	
	\end{tabular}
\end{table}

Ramp events occur when the wind power increases or decreases suddenly in a short duration of time which is typically in the range from 5 minutes to 6 hours \cite{GallegoCastillo2015}.  For a given wind turbine, let us say the ramp threshold power is $\alpha_{th}\%$ of the nominal wind power. Then we can define two ramp thresholds, that is,\vspace{0.1cm}              
\begin{eqnarray}
\Delta P_w^{ramp}&=&\left \{
\begin{aligned}
&+\alpha_{th}\%~~ \text{of}~~ P_{nom} &=& P^{up},\\
&-\alpha_{th}\%~~ \text{of}~~ P_{nom} &=& P^{down},
\end{aligned} 
\right.
\end{eqnarray}
where $P^{up}$ and $P^{down}$ are the upper and lower ramp thresholds respectively depicting ramp-up and ramp-down events in a short period of time. The rated wind speed considered for ramp event study is 12 m/s. Further, in order to forecast the ramp events in the wind speed time series, wind power is calculated for the given sample of data. 

The ramp event is identified when the wind power generated exceeds lower ($P^{down}$) or upper ($P^{up}$) threshold values. To evaluate the forecasting performance, several error metrics such as Root mean squared error (RMSE), normalized mean squared error (NMSE), coefficient of performance ($R^2$) and Theil's U1 and U2 metrics and are given by (\ref{e7})-(\ref*{e11}):
\begin{eqnarray}
\label{e7}
\text{RMSE} &=& \Bigg[\frac{1}{n}\sum_{i=1}^{n}(\hat{s_i}-{s_i})^2\Bigg]^{1/2} \\
\label{e8}
R^2&=&\frac{\sum_{i=1}^{n}(\hat{s_i}-\bar{s})^2}{\sum_{i=1}^{n}({s_i}-\bar{s_i})^2} \\
\label{e9}
\text{NMSE}&=&\frac{\sum_{i=1}^{n}(\hat{s_i}-{s_i})^2}{\sum_{i=1}^{n}({s_i}-\bar{s})^2} \\
\label{e10}
\text{Theil U1}&=&\frac{\sqrt{\frac{1}{n} \sum_{i=1}^{n}\left(\hat{s}_{i}-s_{i}\right)^{2}}}{ \left(\sqrt{\frac{1}{n}  \sum_{i=1}^{n} s_{i}^{2}}+\sqrt{\frac{1}{n} \sum_{i=1}^{n} \hat{s}_{i}^{2}}\right)}\\
\label{e11}
\text{Theil U2}&=&\frac{\sqrt{\frac{1}{n} \sum_{i=1}^{n}\left(\left(s_{i+1}-\hat{s}_{i+1}\right) / s_{i}\right)^{2}}}{\sqrt{\frac{1}{n} \sum_{i=1}^{n}\left(\left(s_{i+1}-\hat{s}_{i}\right) / s_{i}\right)^{2}}} 
\end{eqnarray}  
where $\hat{s_i},s_i,\bar{s}$ are the predicted, actual and mean values of the $n$ testing samples. 


\section{Results and Discussions}
In this Section, we discuss the forecasting performance of the prediction models (Persistence, $\varepsilon$-SVR, LSSVR, TSVR, $\varepsilon$-TSVR, RFR and GBM). The ramp-up and ramp-down events are identified for a threshold of 10\% of nominal wind power. The forecasting process is carried out by splitting the entire dataset into training (80\%) and testing (20\%) sets. For persistence model, the forecast is carried out using two previous dispatch windows. Further, for SVR and its variants, the hyperparameters, that is, RBF bandwidth ($\sigma$) and regularization parameter ($C$) are tuned from the set $[2^{-10},2^{-9},...2^9,2^{10}]$. For random forest regression (RFR) and gradient boosted machines (GBM), the simulations are carried out in R studio using randomForest package. The number of trees used in training phase are 1000 for RFR, while for GBM the learning rate is kept at 0.05 and number of trees are 10000. The input feature set to all the prediction models is a matrix consisting of approximation signal A5 and detail signals (D1,D2, $\dots,$ D5) obtained from wavelet decomposition of wind speed time series. Further, we evaluate, absolute error for ramp-up and ramp-down events as $R^{up}$ and $R^{down}$ respectively.  

For onshore wind farm sites, the datasets labeled are \textbf{a, b, c, d, e} and \textbf{f}. For dataset \textbf{a}, GBM outperforms all the prediction models in terms of RMSE and NMSE. For predicting ramp-up and ramp-down events, TSVR outperforms GBM and RFR. In terms of $R^2$, TSVR gives the best fit close to 1.00 whereas $\varepsilon$-SVR gives a value greater than 1.00 causing over-fitting. The over-fitting in prediction is avoided by using RFR and GBM. 
Similarly, for offshore wind farm datasets (\textbf{A, B, C, D, E \& F}), TSVR outperforms all the models in terms of RMSE and NMSE for predicting wind speed, while for predicting ramp events TSVR gives the best result in terms of $R^2$. Since the datasets are large sized, TSVR, $\varepsilon$-TSVR, RFR and GBM take less CPU time (which is the time spent by the regression method to predict the ramp events) than conventional $\varepsilon$-SVR. The detailed performance metrics for onshore and offshore wind farm sites are depicted in Table \ref{T3} and  \ref{T5}. It is worthwhile to note that, ramp prediction performance for GBM is significantly closer to TSVR in case of offshore wind farms.      

\begin{table*}
		\begin{center}
			\caption{Performance metrics for Onshore wind farm sites}\label{T3}
			\begin{tabular}{|c|c|c|c|c|c|c|c|c|c|}\hline
				Dataset & Model & RMSE &  NMSE & $R^2$& U1 & U2 & $R^{up}$ & $R^{down}$& CPU time \\ 
				& & (m/sec) & & & &  & & & (secs) \\ \hline
				& LSSVR & 0.3602 &  0.0177 &  0.9856 & 0.0024& 0.5678 & 0.0236& NA & 1.46082 \\
				& TSVR & 0.3525 &  0.0177 & 0.9899 & 0.0001& 0.0519&0.0016& NA & 57.41689 \\ 
				\textbf{a}& $\varepsilon$-TSVR & 0.3475 & 0.0178  &  0.9701 & 0.0046& 0.6538 & 0.0036& NA & 7.3946  \\ 
				& RFR & 0.3436 & 0.1251  &  0.9506 & 0.5633& 0.6276 & 0.0789& NA & 4.2693  \\
				& GBM & 0.1197 & 0.0044  &  0.9705 & 0.0115& 0.1539 & 0.0043& NA & 0.7119  \\
				& {Persistance} & 0.4399 & 0.0221  &  1.0145 & 0.0259 & 0.6296& 0.2800& NA & 0.02545  \\
				& $\varepsilon$-SVR & 0.3620 &  0.0178 &  1.0004 & 0.0029& 6.6035 &0.0862 & NA & 238.8274 \\
				\hline
				& LSSVR & 0.0403 &  0.00057 &  0.9748 & 0.0044 & 0.5716 & 0.1169 & 0.1498 & 0.76059 \\
				& TSVR & 0.0268 &  0.0002 & 0.9832 & 0.0029 & 0.444 & 0.0881 & 0.1228 & 23.9192 \\ 
				\textbf{b}& $\varepsilon$-TSVR & 0.0486 & 0.0008  &  0.9758 & 0.0053 & 0.6978 & 0.1098 & 0.1332 & 3.3806  \\ 
				& RFR & 0.06991 & 0.00121  &  0.9730 &  0.0660& 0.7191 & 0.1112 & 0.1421 & 32.11  \\
				& GBM & 0.0651 & 0.0017  &  0.9986 & 0.0078 & 0.0219 & 0.0235 & 0.0307 & 45.93  \\
				& {Persistance} & 0.4399 & 0.0108  &  1.0068 & 0.0190 & 0.6429& 0.2350& 0.2000 & 0.04356  \\
				& $\varepsilon$-SVR & 0.0593 &  0.0013 &  0.9556 & 0.0065 & 0.6517 & 0.1876 & 0.2367 & 84.4812 \\ \hline 
				& LSSVR & 0.1806 &  0.0026 &  0.9263 
				& 0.0141 & 0.2834 & 0.0600 & 0.5072 & 1.41105 \\
				& TSVR & 0.0284 &  0.00002 & 0.9884 & 0.0022 & 0.0034 &0.0014 & 0.1119 & 61.54305 \\ 
				\textbf{c}& $\varepsilon$-TSVR & 0.1964 & 0.0031  & 0.9252 & 0.0154 & 0.5499 & 0.0793 & 0.5588 & 7.82425  \\ 
				& RFR & 1.7407 & 0.2461  &  0.4191 & 0.1478 & 1.3082 & 1.7112 & 5.0778 & 8.3472  \\
				& GBM & 1.2148 & 0.1239  &  0.6060 & 0.1740 & 0.2787 & 0.1661 & 3.4109 & 0.6997  \\
				& {Persistance} & 0.3989 & 0.0129  &  1.0051  & 0.0309 & 0.6207 & 0.3250 & 0.1950 & 0.03783  \\
				& $\varepsilon$-SVR & 0.1937 &  0.0030 &  0.9253  & 0.0152 & 0.4438 & 0.0563 & 0.5227 & 251.86508 \\ \hline
				& LSSVR & 0.0263 &  0.000009 &  0.9927 &  0.0024 & 0.4229 & 0.0300 & 0.0109 & 0.519493 \\
				& TSVR & 0.0040 &  0.00002 & 1.000 & 0.0003 & 0.0662 &0.0004 & 0.00002 & 15.967091 \\ 
				\textbf{d}& $\varepsilon$-TSVR & 0.0732 & 0.0007  &  0.9814 & 0.0067 & 0.6507 & 0.1065 & 0.0471 & 2.067738  \\ 
				& RFR & 0.6014 & 0.0489  &  0.7966 & 0.0543 & 0.9854 & 0.0204 & 0.1416 & 5.187355  \\
				& GBM & 0.2758 & 0.0103  &  0.9585 & 0.0252 & 0.1458 & 0.1001 & 0.1444 & 0.505211  \\
				& {Persistance} & 0.2154 & 0.0063  &  1.0023 & 0.0197 & 0.5417& 0.3550 & 0.3500 & 0.0616  \\
				& $\varepsilon$-SVR & 0.0382 &  0.0001 &  0.9951 & 0.0035 & 0.6014 & 0.0594 & 0.0345 & 33.06520 \\
				\hline
				& LSSVR & 0.0232 &  0.000006 &  0.9935 & 0.0017 & 0.3551 & 0.0626 & 0.0689 & 1.022718 \\
				& TSVR & 0.0042 &  0.00002 & 0.9994 & 0.00004 & 0.004 & 0.0134 & 0.0125 & 43.252339 \\ 
				\textbf{e}& $\varepsilon$-TSVR & 0.0344 & 0.0001  &  0.9921 & 0.0025 & 0.7606 & 0.1058 & 0.1035 & 5.18525  \\ 
				& RFR & 0.09609 & 0.00137  &  0.9607 & 0.0016& 0.2612 & 0.1527 & 0.093129 & 15.87  \\
				& GBM & 0.07739 & 0.0008  &  0.9968 & 0.0059 & 1.3718 & 0.0822 & 0.0160 & 36.72  \\
				& {Persistance} & 0.2517 & 0.0077  &  1.0065 & 0.0185 & 0.6629 & 0.5450 & 0.1700 & 0.002612  \\
				& $\varepsilon$-SVR & 0.0400 &  0.0001 &  0.9974 &  0.0029 & 0.4500 & 0.0603 & 0.0981 & 253.34162 \\
				\hline 
				& LSSVR & 0.0216 &  0.0001 &  0.9971 & 0.0022 & 0.4651 & 0.0475 & 0.1055 & 0.700270 \\
				& TSVR & 0.0030 &  0.000002 & 0.9993 & 0.0003 & 0.0112 &0.0076 & 0.0003 & 17.6919 \\ 
				\textbf{f}& $\varepsilon$-TSVR & 0.0467 & 0.0005  & 0.9922 & 0.0049 & 0.8286 & 0.1660 & 0.0980 & 4.27225  \\ 
				& RFR & 0.085906 & 0.00175  &  0.9552 & 0.0057 & 0.8316 & 0.03032 & 0.08309 & 15.83  \\
				& GBM & 0 .07239 & 0.001219  &  0.9998 &  0.0051 & 0.7216 & 0.01447 & 0.1298 & 44.66 \\
				& {Persistance} & 0.2250 & 0.0118  &  1.0040 & 0.0234 & 0.5833 & 0.3050 & 0.2250 & 0.0494  \\
				& $\varepsilon$-SVR & 0.0440 &  0.0004 &  1.0129 & 0.0046 & 0.7638 & 0.0518 & 0.0776 & 63.81193 \\
				\hline
			\end{tabular}
		\end{center}
	\end{table*}

	\begin{table*} 
		\begin{center}
			\caption{Performance metrics for Offshore wind farm sites}\label{T5}
			\begin{tabular}{|c|c|c|c|c|c|c|c|c|c|}\hline
				Dataset & Model & RMSE &  NMSE & $R^2$& U1 & U2 & $R^{up}$ & $R^{down}$& CPU time \\ 
				& & (m/sec) & & & & &  & & (secs) \\ \hline
				
				& LSSVR &0.0225&	0.00004&	0.9899&	0.0013&	0.352&	0.004&	0.0221&	0.507431\\
				& TSVR&0.0037&	0.00001&	0.9998&	0.0002&	0.0745&	0.0034&	0.0038&	11.28299\\
				\textbf{A}& $\varepsilon$-TSVR&0.0429&	0.0001&	0.9891&	0.0025&	0.5738&	0.0162&	0.0777&	2.0435\\
				& RFR &0.0989&	0.00104&	0.9799&0.0036&	0.5161&	0.1032&	0.05638&	31.2\\
				& GBM&0.0967&	0.0075&	0.9904&	0.0032&	0.5016&	0.1052&	0.0423&	50.59\\
				& Persistance &0.324&	0.0084&	1.0001&0.019&	0.6&	0.185&	0.345&	0.051183 \\
				& $\varepsilon$-SVR &0.0367&	0.0001&	1.0009&	0.0022&	0.5024&	0.0167&	0.0375&	84.006\\
				\hline
				
				&LSSVR &0.0198&	0.00006&	0.9912&	0.0013&	0.3908&	0.008&	0.031&	0.4999\\
				& TSVR&0.0038&	0.000002&	0.9998&	0.0002&	0.0115&	0.0043&	0.0028&	12.5415\\
				\textbf{B}&$\varepsilon$-TSVR &0.0434&	0.0003&	0.9998&	0.0028&	0.7492&	0.1064&	0.0771&	1.9274\\
				& RFR &0.7733&	0.00105&	0.9718&	0.0034&	0.0081&	0.18005&	0.07037&	31.0600\\
				& GBM &0.06248&	0.00068&	0.9994&	0.003&	0.0071&	0.180035&	0.07988&	44.5800\\
				& Persistance&0.1655&	0.0048&	1.0067&	0.0106&	0.6389&	0.38&	0.3&	0.1496\\
				& $\varepsilon$-SVR&0.0367&	0.0002&	1.0001&	0.0024&	0.0074&	0.0186&	0.0849&	27.9990\\
				\hline
				
				&LSSVR&0.0232&	0.00006&	0.9944&	0.0021&	0.3838&	0.0217&	0.0305&	0.514049\\
				& TSVR &0.0064&	0.000004&	0.9993&	0.0005&	1.0497&	0.0001&	0.0012&	13.034972\\
				\textbf{C}& $\varepsilon$-TSVR&0.0463&	0.0002&	0.9871&	0.0043&	1.0083&	0.056&	0.076&	1.90010\\
				&RFR&0.07823&	0.00724&	0.9763& 0.0051&	1.0091&	0.7805&	0.0896&	30.58\\
				& GBM&0.07146&	0.0039&	0.9863&	0.0041&	1.0012&	0.1687&	0.04527&	43.47\\
				&Persistance&0.2577&	0.0079&	1.0015&	0.0238&	0.5&	0.55&	0.555&	0.04261\\
				&$\varepsilon$-SVR&0.0626&	0.0004&	0.9875&	0.0058&	1.0347&	0.0765&	0.0866&	39.806\\ \hline
				&LSSVR&0.0244&	0.000004&	0.9987&0.0014&	0.057&	0.0139&	0.0119&	0.5023\\
				&TSVR&0.0032&	0.00007&	0.9999&		0.0001&	0.0213&	0.0024&	0.0027&	12.7748\\
				\textbf{D}&$\varepsilon$-TSVR&0.0352&	0.00009&	0.9994&		0.002&	1.5215&	0.0338&	0.0229&	1.9140\\
				&RFR&0.09762&	0.00072&	0.9758&	0.0042&	0.7166&	0.09793&	0.04394&	38.7200\\
				&GBM &0.0793&	0.001004&	0.9868&	0.0056&	0.8166&	0.00686&	0.0029315&	55.4600 \\
				&Persistance&0.4252&	0.0138&	1.0038&	0.0236&	0.75&	0.055&	0.45&	0.0492\\
				&$\varepsilon$-SVR&0.0379&	0.0001&	1.0105&	0.0021&	0.6116&	0.0136&	0.0123&	80.1100\\ \hline
				
				&LSSVR&0.0263&	0.00005&	0.9994&	0.0015&	0.4724&	0.0032&	0.0626&	0.50647\\
				&TSVR&0.004&	0.00001&0.9997&0.0002&	0.0309&	0.0004&	0.0018&	12.743\\
				\textbf{E}&$\varepsilon$-TSVR&0.0437&	0.0001&	0.9994&	0.0024&	0.633&	0.004&	0.1313&	1.90508\\
				&RFR&0.1059&	0.0008&	0.9761&0.0059&	0.8372&	0.2281&	0.3498&	37.43\\
				&GBM&0.08278&	0.0004&	0.9951&0.0096&	1.8164&	0.08923&	0.0112&	57.2\\
				&Persistance&0.3866&	0.0109&	1.0057&	0.0215&	0.6&	0.19&	0.105&	0.052525\\
				&$\varepsilon$-SVR&0.0338&	0.00008&	1.0041&		0.0019&	0.1171&	0.0885&	0.0112&	90.5178\\
				\hline
				
				&LSSVR&0.0206&	0.00003&0.9991&	0.0012&	0.3587&	0.0015&	0.044&	0.58391\\
				&TSVR&0.0029&	0.000006&	0.9995&	0.0001&	0.173&	0.0003&	0.0065&	12.78991\\
				\textbf{F}&$\varepsilon$-TSVR&0.037&	0.0001&	0.9912&0.0022&	0.419&	0.0034&	0.0408&	1.87901\\
				&RFR&0.1059&	0.00084&	0.9774&0.0062&	1.2043&0.1264& 0.2182&	33.01\\
				&GBM&0.0929&	0.00064&	0.9976& 0.0054&	0.3996&	0.04356&	0.07129&	48.38\\
				&Persistance&0.3244&	0.0079&	1.0036&0.0139&	0.5&	0.21&	0.405&	0.05162\\
				&$\varepsilon$-SVR&0.033&	0.00008&	1.0001&	0.0019&	35.36&	0.0358&	0.0535&87.76891\\
				\hline
			\end{tabular}
		\end{center}
	\end{table*}

Results obtained in aforementioned section depict that SVR based regressors have an upper hand over RFR and GBM based prediction methods. While it is worthy to note that, to study the ramp events the time interval between consecutive samples plays an important role in deciding the threshold value. In a study carried out by Ouyang et al., the wind speed data collected is sampled every 15 minute for a wind farm site in China and ARIMA model is used to forecast wind power \cite{Ouyang2019}. 

Further, swinging door algorithm is used to detect ramp events. The RMSE values for the ramp event prediction are found to be in range of 32\%. Using machine learning techniques such as classical SVR, GPR, MLP and ELM, the ramp events are predicted for time interval 6hr and RMSE values are found to be in range of (5-7 MW). However, wind speed time-series of onshore and offshore wind farm has a lot of variability in terms of magnitude. Thus, a ramp event study with 10 minute as time interval poses critical impositions in market clearing and day-ahead scenarios. 

The current work deals with predicting wind power ramp events with 10 minute sampling interval for onshore and offshore wind sites. With offshore wind farms, the wind speed being high and variable, the probability of ramp event increases. Thus, in the current work, we extend the ramp event prediction study by incorporating advanced machine learning algorithms like variants of SVR, random forest regression and gradient boosted machines. Results reveal that TSVR based prediction model yield lowest error for ramp-up and ramp-down events. 

\subsection*{Uncertainties in Ramp events}
Ramp event in onshore and offshore wind farms is a challenging issue which could 
adversely affect the associated power systems. 
The randomness in wind speed stimulated by turbulent air flow can cause unwanted vibrations in turbine blade and tower thus questioning its structural stability. Randomness in a signal or time-series can often be expressed by calculating the log energy entropy. In order to critically examine a ramp event signal, expressed as (\ref{e1}), we decompose the obtained signal by two signal processing techniques such as wavelet transform decomposition (WT) and empirical model decomposition (EMD). While wavelet transform decomposes a signal into low-frequency and high-frequency components, EMD on the other hand is capable of extracting oscillatory features from the signal for a dominant frequency. The ramp signal is decomposed using db4 wavelet filter using wavelet transform and into 5 intrinsic mode functions (IMFs) using EMD. 

\begin{figure}[H]
	\centering
	\hspace*{-0.3cm}\includegraphics[width=1.2\linewidth]{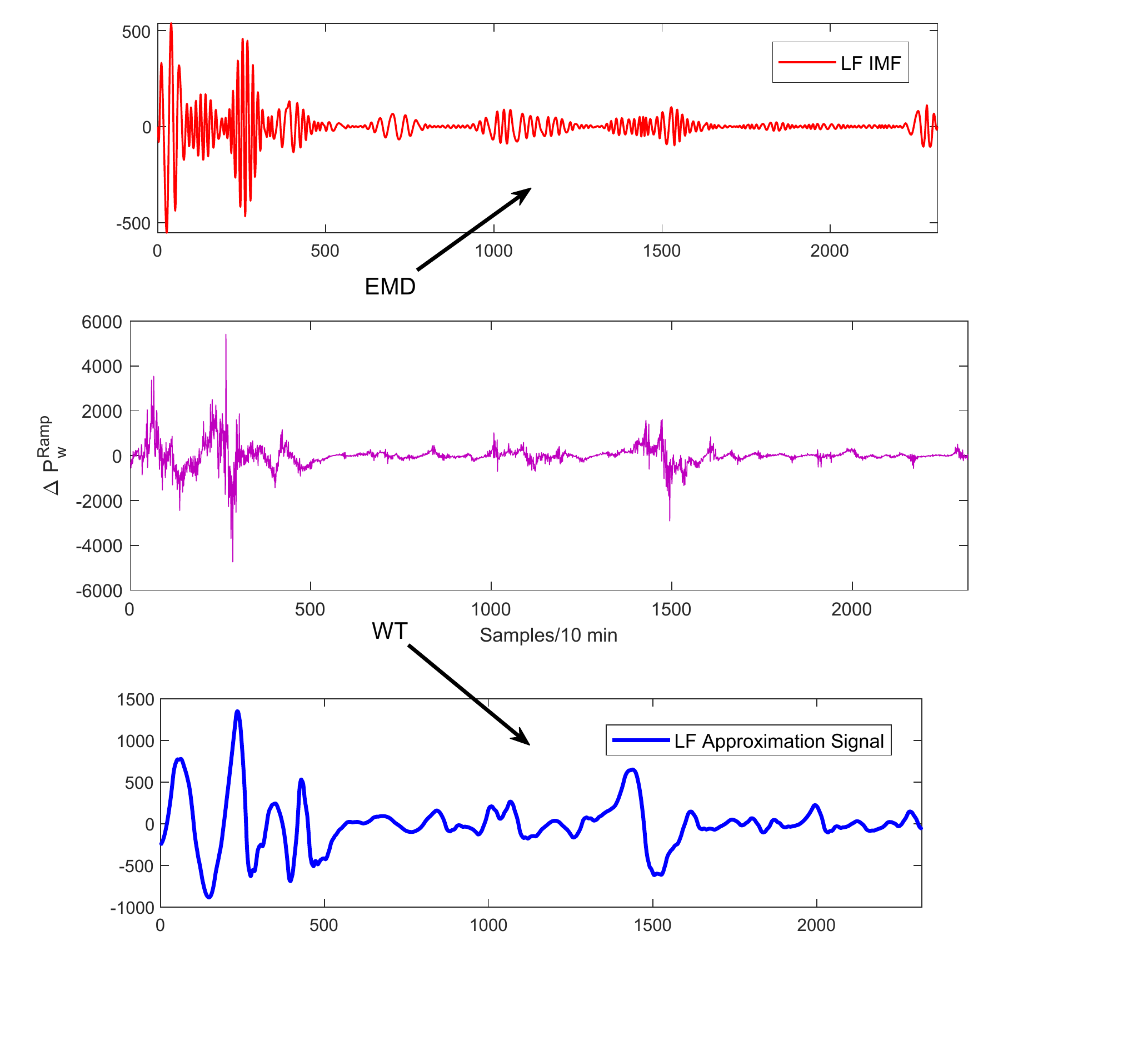}
	\vspace{-1.3cm}
	\caption{Ramp event signal and its decomposition using wavelet transform and EMD}
	\label{nokia}
\end{figure}

The log energy entropy for a signal $h(t)$ is given as
\begin{eqnarray}
E\{h(t)\}=\sum_{t=0}^{T}\log(h(t)^2).
\end{eqnarray} 
A typical ramp event signal and its signal decomposition using wavelet transform and EMD is illustrated in Figure \ref{nokia} which shows the low frequency (LF) signals obtained from wavelet transform and EMD and are used to assess the randomness in the ramp event signal by calculating the log energy entropy of predicted ramp signals based on TSVR, RFR and GBM are depicted in Table \ref{T8}. 

\begin{table}[H]
	\caption{Log energy entropy for ramp event signals}\label{T8}\vspace{0.2cm}
	\centering
	\begin{tabular}{|c|c|c|c|}\hline
		\multirow{1}{*}{Dataset} & \multirow{1}{*}{Model} & \multicolumn{2}{c|}{Log energy entropy}\\ \cline{3-4}
		& & WT & EMD \\ \hline
		\textbf{a} & TSVR &  6.6094$\times$10$^3$ & 5.6985$\times$10$^3$ \\
		& RFR  &  6.2866$\times$10$^3$ & 5.5948$\times$10$^3$\\
		& GBM  &  6.7926$\times$10$^3$ & 5.9943$\times$10$^3$ \\ \hline
		\textbf{A} & TSVR &  1.7259$\times$10$^4$ & 1.5646$\times$10$^4$ \\
		& RFR  &  2.0676$\times$10$^4$ & 1.2297$\times$10$^4$\\
		& GBM  &  2.0776$\times$10$^4$ & 1.4621$\times$10$^4$ \\ \hline
	\end{tabular}
\end{table}
We observe that for a particular prediction model, the log energy entropy based on WT is more than that of EMD suggesting higher order randomness in ramp event signal. As far as uncertainties are concerned, the decomposition based prediction models yield better ramp prediction than single methods.

\section{Conclusion}
In the current work, wind power ramp events are studied for 6 onshore and 6 offshore wind farm sites. Wind speed data for the month of March 2019 with 10 minute sampling interval is collected. The available data is transformed to a hub height of 90m for all the datasets. A threshold value for change in wind power is chosen as 10\% of nominal wind power and the ramp event points are assessed for absolute error. Machine learning based prediction models are compared with benchmark persistence model. TSVR based model gives minimum absolute error for both, ramp-up and ramp-down events. The coefficient of performance ($R^2$) for RFR and GBM in case of offshore wind farm datasets is found to be close to 1.00 which indicates a good agreement between predicted and actual ramp values. Further, it is observed that error in predicting ramp-down event is more than in ramp-up event. Overall, machine learning based prediction models are found to be good approximators for analyzing 10-min wind ramps. The randomness in wind power ramp series in case of TSVR, RFR and GBM is evaluated using signal decomposition techniques and with RFR the randomness is found to be minimum. Thus, an accurate machine intelligent model like TSVR and GBM in this case, affirms a stable grid operation in presence of large ramp-up and ramp-down events.

\bibliographystyle{IEEEtranTIE}
\bibliography{tonystark}\ 

\end{document}